\documentstyle[aps,preprint,epsbox]{revtex}

\begin{document}
\draft
\preprint{}
\title{Ordered phase and phase transitions\\
in the three-dimensional generalized six-state clock model}

\author{
Norikazu Todoroki, Yohtaro Ueno and Seiji Miyashita
}
\address{
Department of Applied Physics, University of Tokyo, Bunkyo-ku, Hongo, Tokyo 113-8656\\
Department of Physics, Tokyo Institute of Technology,
Oh-okayama, Meguro, Tokyo 152-8551, Japan
}
\date{\today}
\maketitle
\begin{abstract}
We study the three-dimensional generalized six-state clock model at values of the energy parameters, at which the system is considered to have the same behavior 
as the stacked triangular antiferromagnetic Ising model and 
the three-state antiferromagnetic Potts model.
First, we investigate ordered phases by using the Monte Carlo twist method (MCTM). 
We confirmed the existence of an incompletely ordered phase (IOP1) at intermediate temperature, besides the completely ordered phase (COP) at low-temperature.
In this intermediate phase, two neighboring states of the six-state model mix, while one of them is selected in the low temperature phase.
We examine the fluctuation the mixing rate of the two states in IOP1 and clarify that the mixing rate is very stable around 1:1. 

The high temperature phase transition is investigated by using 
non-equilibrium relaxation method (NERM). We estimate the critical exponents $\beta=0.34(1)$ and $\nu=0.66(4)$. 
These values are consistent 
with the 3D-XY universality class. The low temperature phase 
transition is found to be of first-order by using 
MCTM and the finite-size-scaling analysis. 
\end{abstract}

\section{INTRODUCTION}

The systems of Z$_{6}$ symmetry in the three dimensions (3D)
have attracted interests because of their peculiar nature of orderings.
In particular, there have been many works on the nature of intermediate temperature phases.

In two dimensions, the six state clock model shows three phases. 
At intermediate temperature, the discreteness of the six states 
is irrelevant and the system shows the same properties as that of XY model,
which is the Kosterlitz-Thouless phase where the long range order does 
not exist but the correlation length of the order diverges \cite{intro0a,intro0b}. 

In three dimensions, the six state clock model also shown 
natures which are similar to the ordered phase of the XY model.
However it found that there exists only a phase where
one of the six state of the model is chosen as the 
long range order \cite{intro5}.
The apparent behavior is due to the large fluctuation 
at the higher temperature region of the ordered phase.
However, if we generalize the energy structure of the model,
a new type of intermediate phases appears.

Ueno {\it et al} have introduced three-dimensional generalized six-state clock model
 (3D-6GCL model)\cite{intro3}. 
The 3D-6GCL model can be regarded as prototypes of wide range of Z$_{6}$ models.
That is, it represents various categories of the Z$_{6}$ models
 according to the values of the energy parameters of the model.
The existence of intermediate phase have attracted interest for the following models:
 three-dimensional three-state antiferromagnetic
Potts model (3D-3AFP model)\cite{intro15}, 
and stacked triangular antiferromagnetic Ising model (STAFI model)\cite{intro16}. 
Although, the existence of various type of intermediate phases for these models 
have been proposed, there is no general understanding on the nature of the intermediate phase(s) 
\cite{intro5,intro3,intro15,intro16,intro6,intro7,intro10,intro8,intro9,intro12}.
If we consider only one universality class for models with Z$_6$ symmetry, 
we conclude that the intermediate phase is an apparent phase as we found in the regular six state model.
However, it has been clarified that the situation is not so simple and interesting new type of 
phase exists in the generalized case. 
In fact, the 3D-6GCL model was introduced in order to understand general properties of other Z$_{6}$
models \cite{intro3}. In the present paper, we study the 3D-6GCL model with a set of the parameters which 
is considered to correspond to the STAFI model and the 3D-3AFP model.

The Hamiltonian of the 3D-6GCL model is given by
\begin{eqnarray}
{\cal H}=\sum_{\langle i,j\rangle }\varepsilon_{0}\delta_{n_in_j}
+\varepsilon_{1}\delta_{n_in_j\pm 1}
+\varepsilon_{2}\delta_{n_in_j\pm 2}
+\varepsilon_{3}\delta_{n_in_j\pm 3},
\end{eqnarray}
where $\langle i,j\rangle$ runs over nearest neighbor pairs, 
and $n_i$ is the spin variable which takes one of $1, 2, \cdots$ and $6$.
We set $\varepsilon_0=0\ge\varepsilon_1\ge\varepsilon_2\ge\varepsilon_3$ as in Fig.\ref{6GCL-1}.
Namely, we only study ferromagnet cases 
where all the spins occupy the same state in the ground state. 
We expect that this ferromagnetic ordered phase appears at the low temperature.
This ordered phase is called completely ordered phase (COP). 
Since the COP has six degenerate states, the order parameter space of this model
 is illustrated by a hexagon as in Fig.\ref{6GCL-2}. The bold solid line 
represents the one of the six COP states. The dotted line represents
 a mixture of neighboring two COP states which is a candidate of an intermediate phase, 
which we call incompletely ordered phase (IOP1).
As to the distribution of the spin state in the incompletely ordered phase, 
there have been proposed the two kinds depicted in Fig.\ref{IOP}. 
The IOP1 discussed above corresponds to Fig.\ref{IOP}(a). 
The type of Fig.\ref{IOP}(b) was called IOP2 which was one of the possible mixed state. 
For $\varepsilon_{1}= 0$, 
the COP does not appear because the ground state is macroscopically degenerate. 
This degenerate state is IOP1 and corresponds to the ground state of the 3AFP model. Taking the approach of physical percolation to phase transitions, Ueno rigorously proved that for $\varepsilon_{1}=0$, 
at least one kind of IOPs can exist at $T<\varepsilon_{2}=\varepsilon_{3}$ \cite{intro4}. 
For $\varepsilon_{1}>0$ 
, Ueno and Kasono proposed that this model has
two incompletely ordered phases (IOP1, IOP2) besides the COP by using the Monte Carlo twist method (MCTM) \cite{intro3}. 
In this paper, we reexamine the nature of the intermediate phase(s) in detail.

On the 3D-3AFP model, so far various types of phases 
at intermediate temperatures have been proposed in the previous studies, 
such as the permutationally symmetric sublattice phase (PSS phase) which corresponds the IOP2 \cite{intro6,intro7} or the 
gapless phase which is similar to the ordered phase of the three dimensional XY model
which we call XY phase \cite{intro10}.
Recent theoretical and numerical studies on the 3D-3AFP model have revealed that 
the existence of these phases
are fake owing to the finite size effect \cite{intro1,intro2,intro14}.
There is only a low-temperature phase that corresponds to the IOP1.

We will show that the correlation length of the fluctuation of 
the order parameter in IOP1 is finite. 
However the correlation length just below the critical point is so long that 
the phase looks like the ordered phase of the 
3D XY model where the correlation length is infinite although the long range order exists. 
In order to distinguish the difference between 
the XY phase and IOP1 we have to be careful in the examination of properties of the phase.

In this paper, we study a 3D-6GCL model by using
 the Monte Carlo (MC) simulation and 
clarify properties of the phase at intermediate temperature and natures of the phase transitions.
In this study, we choose the energy parameters of the 3D-6GCL model 
as $\varepsilon_{0}=0$, $\varepsilon_{1}=0.1$ 
and $\varepsilon_{2}=\varepsilon_{3}=1.0$ which are the same as those in \cite{intro3}. 

In order to examine the possibility of many ordered phases by the MCTM,
one needs a full set of appropriate boundary conditions
which cause different types of domain-walls, 
and calculate size dependence of the excess free energy.
In the present study, we deny the existence of IOP2 as intermediate phase 
from the dependence of the excess free energy on the boundary conditions.
The intermediate phase shows apparent property of the XY phase where Z$_6$ discreteness is irrelevant.
However, we also deny the existence of XY phase in the thermodynamic limit. 
This apparent property is attributed to the long correlation length
 at the high-temperature region of the intermediate phase. 
Finally, we conclude that only IOP1 exists as the intermediate temperature. 
Appearance of the fake XY phase near the critical point has been also found in 
the six-state clock model,
where the Z$_6$ discreteness is very weak
in the ordered phase near the critical point.
However the Z$_6$ discreteness is always relevant in the ordered phase \cite{intro5}.

Moreover, we examine the properties of the phase transitions. We consider that the high temperature phase transition belongs to 3D-XY universality class.
In order to confirm whether this transition belongs to 3D-XY universality class, 
we study this transition by using non-equilibrium relaxation method (NERM) and estimate 
the critical temperature and critical exponents. 
Ueno and Kasono pointed out that the low-temperature transition between the IOP1
and the COP is of the first order because the two phases have no symmetry relation of group and subgroup\cite{intro3}. We investigate this phase transition by using MCTM and the usual finite size scaling analysis, and confirmed that it is of the first order.

The outline of this paper is as follows. 
In Sec.II, we study properties of the intermediate phase. 
In Sec.III, we study that the high temperature transition and
in Sec.IV, we investigate the low temperature phase transition, and the Sec.V is for summary and discussion. 

\section{INTERMEDIATE PHASE}
The properties of the phases of the 3D-6GCL model have been studied by the MCTM \cite{intro3},
which is a simulation method under special boundary conditions to detect the domain-wall excess free energy. 
We prepare systems with fixed boundaries in one direction 
and impose a periodic boundary condition in the other directions. 
As a reference system, we set the same state $\alpha$ at the boundaries ($\alpha$-$\alpha$) 
in which we expect that the ordered phase $\alpha$ appears. 
We also prepare a system where we set the states $\alpha$
 and $\beta$ at the boundaries ($\alpha$-$\beta$) in which we expect an interface.
The interfacial free energy is defined as the excess free energy between the two systems:
\begin{eqnarray}
\Delta F^{\alpha\beta}(T,L)=F^{\alpha\beta}(T,L)-F^{\alpha\alpha}(T,L),
\end{eqnarray}
where $L$ is a liner size of the system. Here we consider an $L\times L\times  L$ system. 
The size dependence of $\Delta F^{\alpha\beta}(T,L)$ is given by the following asymptotic
behavior for large $L$
\begin{eqnarray}
\Delta F(T,L)\sim A(\alpha,\beta)L^{\psi_{\alpha\beta}(T)},
\end{eqnarray}
where $\psi_{\alpha\beta}(T)$ is called stiffness exponent. 
The ordered phase is classified by the value of stiffness exponent as 
\begin{eqnarray}
\psi_{\alpha\beta}(T)=
\left \{
\begin{array}{l}
D-1 \hspace{5mm}\mbox{for a domain-wall type interface ($\xi < \infty$)}\\
D-2 \hspace{5mm}\mbox{for a gepless interface (spin-wave type $\xi =\infty$)}\\
\mbox{non-integer for a new type on interface}\\
\end{array}
\right. .
\end{eqnarray}
When the interface does not appeare, the stiffness exponent becomes negative 
because $\Delta F(T,L)\rightarrow 0$ for $L\rightarrow\infty$. 

In the present work, we use three sets of boundary conditions which are shown in Fig.\ref{BC}. 
There are two kinds of boundary conditions; 
in $\phi_{1}$ we fix the boundaries to be two of the low temperature phases. The relative angle (twist angle) between the two phases can be $\frac{\pi}{3}$, $\frac{2\pi}{3}$ and $\pi$.
Here we take the cases $\frac{\pi}{3}$ and $\pi$ which are shown in Fig.\ref{BC} (a) and \ref{BC} (b), respectively. 
In $\phi_{2}$ we fix the boundaries to be two of the IOP1s in Fig.\ref{BC} (c) (twist angle $\pi/3$). 
As shown in Table \ref{IPCMCTMT1}, we distinguish the five phases from the signs and the values of the stiffness exponent.
We perform MC simulation for the lattices with liner size $L=8, 10$, and 12 with boundary conditions $\phi_{1}(\pi)$, $\phi_{1}(\pi/3)$ and $\phi_{2}(\pi/3)$. At each temperature, after discarding first $5000L$, $5000L$ and $10000L$ steps, we calculate quantities
of interest using data of next $25000L$, $25000L$ and $50000L$ steps for the three sets of boundary conditions, respectively.

In Fig.\ref{IPC_PSI_3BC}, we show the temperature dependence of $\psi$ under the boundary conditions. 
From this figure, we may conclude the existence of the disorder phase, the XY phase, the IOP1 and the COP, 
as the temperature decreases. 
The high temperature transition point is estimated to be $T_{\mbox{C}}\sim 1.35(5)$, 
and the low temperature transition point is to be $T_{\mbox{F}}\sim 0.30(5)$. 
We see a change around $T\simeq 0.7$ which can be a phase transition between the XY phase and IOP1.

In order to confirm the region of phases, we investigate the size dependence of the stiffness exponents for 
$\phi_{1}(\pi /3)$ for larger sizes. The region of the XY phase decreases as 
$L$ becomes large (Fig.\ref{IPC_PSI_P3SD}).
The size-dependence strongly suggests 
that the region of the XY phase disappears in the limit $L\rightarrow\infty$  
and only the IOP1 exists in the intermediate temperature region.

If we consider the domain wall between the ordered states of the type of IOP1,
there is no domain wall in the $\phi_1 (\pi/3)$. On the other hand, one domain wall appears in the case of $\phi_2 (\pi/3)$, 
and two domain walls appear in the $\phi_1 (\pi)$. 
Therefore, stiffness exponent for the cases $\phi_1 (\pi)$ and $\phi_2 (\pi/3)$ should be the same.
However, we notice that the difference of 
the values of stiffness exponents of $\phi_{1}(\pi)$ and $\phi_{2}(\pi/3)$ is significant at the IOP1 as we see at Fig.\ref{IPC_PSI_3BC}. 
The reason for this difference is that the effect of twisting in the system by $\phi_{1}(\pi)$ is stronger 
than one of the twisting by $\phi_{2}(\pi/3)$. 
Actually, when we examine $\psi_{1}(\pi)$ and $\psi_{2}(\pi /3)$ in much larger lattice sizes and 
we confirm that the values of 
the stiffness exponents $\psi_{1}(\pi)$ and $\psi_{2}(\pi/3)$ approach the same value 
$\psi=2$ which indicates the domain-wall type order. 
In Fig.\ref{profile}, we show a snapshot of the domain-wall of the boundary in $\phi_2(\pi/3)$ 
of $L=160$ at $T=0.4$, where we find a localized domain-wall clearly.

Next, following the idea of Oshikawa \cite{intro1} for the 3D-3AFP model, 
we make the finite size scaling plots of $-\langle\cos 6\theta\rangle$ 
by ordinary Monte Carlo method on lattices $L=12, 18, 24$ and $30$. 
Here, $-\langle\cos 6\theta\rangle$ is the order parameter for the Z$_6$ symmetry breaking 
and $\theta$ is the angle of magnetization from one of the COP states.
 After discarding first $10000L$ steps, we calculate quantities of interest for next $40000L$ steps at each temperature. 
In Fig.\ref{IPC_P6_SC}, we show the scaling plots 
in the form of the scaling function given by Oshikawa
\begin{eqnarray}
\langle \cos 6\theta\rangle\sim f(cL^2(T_C-T)^{\nu |y_6|}).
\end{eqnarray}
Here, we put the value of the critical temperature $T_{\mbox{C}}=1.384$ which was obtained
by using the NERM which will be explain a later. We obtain a good scaling plot when 
we use $\nu |y_{6}|=3.2$. 
It is important to notice here that behavior of the order in the region of the IOP1 is expressed 
by a scaling function using the high temperature critical point $T_{\mbox{C}}$. This fact indicates 
that other phases do not exist between the IOP1 and the disordered phase
because the fluctuation in the IOP1 phase is controlled by the high temperature critical point. 
However, it is very difficult 
to estimate the accurate value of this exponent because 
the strong finite size effect appears 
in the intermediate phase even at the largest system which we can calculate.
After all, we conclude that the 3D-6GCL model with
 $\varepsilon_{0}=0$, $\varepsilon_{1}=0.1$ and $\varepsilon_{2}=\varepsilon_{3}=1.0$ 
has two phases, i.e. IOP1 as the intermediate phase and COP as the low temperature phase.

In order to investigate the nature of IOP1, we examine the following quantity;
\begin{eqnarray}
\chi =\frac{1}{L^{3}}(\langle (p_1-p_2)^2\rangle -\langle (p_1-p_2)\rangle^2),
\end{eqnarray}
where $p_i$ denotes the number of the spin variables which take the value of the $i$-th state. 
We adopt the boundary conditions $\phi_2(0)$.
The asymptotical form of $\chi$ is written $\chi\sim AL^x$.
The exponent $x$ depends on the type of the order.
When the fluctuation is of the XY phase, we expect $x=2$.
On the other hand, when the fluctuation has a finite correlation length, $x=0$.

We perform MC simulations for the lattices with the liner sizes $L=8, 10, \cdots$ and $24$. 
For each temperature, after discarding first $10000L$ steps, we calculate $\chi$
 using data of next $50000L$ steps.
In Fig.\ref{YOKOYURA}, we show the size and temperature dependence of $x$. 
We find that the exponent $x$ decreases in the IOP1 region. Moreover, we perform MC simulation for the lattice up to the liner size $L=80$ at $T=0.4$ (in the IOP1 region).
In Fig.\ref{YOKOYURA2}, we show the temperature dependence of $\chi$ at $T=0.4$. The value of $\chi$ approaches the constant as $L$ increases. 
Thus we consider that the value of $x$ decreases and approaches 0 as $L\rightarrow \infty$, and
we conclude that the fluctuation in IOP1 is the type of the domain-wall and the correlation length of the fluctuation of the order parameter
is finite. 

\section{HIGH TEMPERATURE PHASE TRANSITION}
The high temperature phase transition 
in Z$_{6}$ models belongs to the 3D-XY universality class \cite{intro1}. 
Indeed, the transitions in the 3D-3AFP model \cite{intro2} 
and the six-state clock model \cite{intro5} are found to belong to the 3D-XY universality class. 
We consider that the high temperature phase transition in the 3D-6GCL model also belongs to 3D-XY universality class. 

We investigate this phase transition by the NERM \cite{HPT1,HPT2,HPT3}. 
The NERM is an efficient numerical technique to estimate the critical point and critical exponents from a dynamical
process toward the equilibrium state from an ordered state.
The decay of the order parameter $m(t)$ shows a power-low only at the critical point.
Detecting such point, a precise determination of the critical temperature can be done.
The asymptotical form of $m(t)$ at the critical temperature is written as
\begin{eqnarray}
m(t)\sim t^{-\lambda_m}.
\end{eqnarray}
To examine asymptotic behavior of $m(t)$ clearly, we introduce a local exponent $\lambda_m (t)$
\begin{eqnarray}
\label{NER1}
\lambda_m(t)\equiv -\frac{d\log m(t)}{d\log t}.
\end{eqnarray}
This exponent $\lambda_m(t)$ corresponds to $\beta/z\nu$. 
Further, we consider the following function and the local exponent.
\begin{eqnarray}
\label{NER2}
f_{mm}(t)\equiv\left [ \frac{\langle m^2\rangle}{\langle m\rangle^2 }-1 \right ], \hspace{1cm}\lambda_{mm}\equiv \frac{d\log f_{mm}(t)}{d\log t},
\end{eqnarray}
where $\lambda_{mm}(t)$ corresponds to $d/z$.
Therefore, we obtain the exponents $\beta/\nu$ and $z$ independently from these quantities.
We simulate the relaxation process starting from an initial state which is set to be 
the IOP1 state, and measure the magnetization $m(t)$.
In Fig.\ref{Hptc_lm}, we show $\lambda_{m}(t)$ for the temperatures near the $T_{\mbox{C}}$.
Here, MC simulation is performed in a lattice $L=60$. We show the MC steps and the number of samples in Table \ref{Hptc_det}.
The curve for $T=1.385$ turns up while the curve for $T=1.383$ turns down. 
Therefore we locate the transition temperature in $1.383 < T_{\mbox{C}} < 1.385$, denoting $T_{\mbox{C}}=1.384(1)$. 

At this accurate value of $T_{\mbox{C}}$, we calculate the relaxation 
of quantities. We perform about 160000 independent runs up to 200MCS
and average the process to obtain $m(t)$.
We obtain the critical exponents $\beta/\nu$ and $z$
by making use of (\ref{NER1}) and (\ref{NER2}). 
In Figs.\ref{Hptc_Z} 
and \ref{Hptc_BPN}, we show $z(t)$ and $\beta/\nu(t)$, respectively. 
From the extrapolated values of $z(t)$ and $\beta/\nu(t)$, to $1/t=0$, 
we estimate $z=2.05(5)$ and $\beta/\nu=0.515(10)$. 

Moreover, assuming $T_{\mbox{C}}=1.384(1)$ and $\beta/\nu=0.515(10)$, we estimate $\beta$ 
from the scaling plots of data of the magnetization which are obtained from the ordinary equilibrium MC simulation. 
Seeking the value of $\beta$ by which the date collapse into a scaling function, 
 we estimate $\beta =0.34(1)$ for the best fit (Fig.\ref{Hptc_Beta}). Since the values, $\beta=0.34(1)$ 
and $\nu=0.66(4)$, are close to the 3D-XY universality class ($\beta=0.345$, $\nu=0.669$), we conclude that this phase transition belongs to the 3D-XY universality class.

\section{LOW TEMPERATURE PHASE TRANSITION}
While many researchers have studied the high temperature transition in the
Z$_{6}$ model, but few study has been done for
the low temperature one. Ueno and Kasono pointed
out that the low temperature transition between the IOP1 and the COP is
of first-order because they have no symmetry relation between group and
subgroup which is usually seen in the second-order transition. 

In this paper we study this problem by a finite-size-scaling analysis 
of $\langle\cos 6\theta\rangle$.
If the transition is of first order, the order parameter is scaled 
by $\langle\cos 6\theta\rangle\sim f(\epsilon L^{d})$ where $\epsilon=(T-T_{\mbox{C}})/T_{\mbox{C}}$, 
$d$ is the dimension of the system and $f(x)$ is a finite size scaling function. 
We obtain a good scaling plot at $T_{\mbox{F}}=0.295(5)$ depicted Fig.\ref{LPTC_P6_SC}. 
Thus we confirm this phase transition is of first-order.

We also confirm this result by an analysis of extension of the MCTM \cite{LPT1,LPT2}.
Now, we study the interfacial energy $\Delta E$ 
is defined in the same way for the excess energy due to the boundary condition causing the interface, eq.(2).
In a completely ordered phase, the size dependence of the excess energy becomes $\Delta E\sim L^{d-1}$.
At the critical point, it becomes $\Delta E\sim AL^{1/\nu}$.
We define the stiffness exponent of the energy $\psi_E$ as $\Delta E\sim AL^{\psi_E}$.
 For the 3D Ising model at the critical point $\psi_E=1.59$ which is considered to be a large value among the 3D models exhibiting the second order phase transition. 
We study $\psi_E$  for the present phase transition. 
We use the boundary condition $\phi_1(\pi/3)$. 
In Fig.\ref{LPTC_MCTM}, we show the temperature and the size dependence of $\psi_{\mbox{E}}$.
We confirm that $\psi_{\mbox{E}}$ approach to 3 which is much larger than the values of $\psi_E$ of the second order phase transition in 3D models.
For the first order phase transition, it is known that $\psi_E=D$ from the argument of wetting\cite{LPT2}. 
This result also suggests that the present phase transition is of first order.

\section{SUMMARY and DISCUSSION}
We study successive phase transitions 
in the 3D-6GCL model and found an intermediate phase and two phase transitions.
First, it has been revealed that the intermediate phase is 
single phase of the IOP1. In the high temperature region of the intermediate
phase, the correlation length is very large and the system shows an apparent XY behavior. 
In this region the stiffness constant is very sensitive to the boundary condition. 
Therefore, the wrong results were obtained in the previous numerical studies \cite{intro3}. 
The present results agree with the phase diagram of the 3D-3AFP model obtained by Kishi and Ueno,
and our results of the STAFI model \cite{DIS1}. 
Thus, we have obtained the same intermediate phase in the 
three 3D Z$_{6}$ models, i.e. the 3D-6GCL model, the 3D-3AFP model and STAFI model. 
Moreover, we found that the mixing rate of the two states 
in the intermediate phase is steady and does not show anomaly fluctuation 
of the rate proposed preciously \cite{intro3}.
This IOP1 with the steady mixing of two COP states can be regarded as 
partially disordered phase proposed by Mekata\cite{mekata} in the original work on the 
triangular antiferromagnets. 
Here we understand that the entropy effect due to the flustration allows different type of intermediate phases
in two and three dimensional model.

Secondly, we examine properties of the high temperature phase transition,
and the transition temperature is estimated to be $T_{\mbox{C}}=1.384(1)$ with 
the critical exponents $\beta =0.34(1)$, $\nu =0.66(4)$ and $z=2.05(5)$. 
These values of the exponents are close to those of 3D-XY universality class. 
We conclude that the universality class of this transition belongs to the 
3D-XY universality class. 

Thirdly, we investigate the low temperature phase transition in the 3D-6GCL 
model by using the finite-size-scaling analysis and the MCTM. 
All of the results support that this transition is 
of first-order. This first-order phase transition is a transition 
at which the $\langle\cos6\theta\rangle $ changes from $-1$ 
to $1$ discontinuity. 

\section*{ACKNOWLEDGMENTS}
We would like to thank F. Matsubara, M. Oshikawa, N.Ito and Y. Ozeki for useful discussions. The numerical calculations were performed on the supercomputers in the Computer Center of the University of Tokyo.


\begin{table}
\caption{Values and signs of stiffness exponents for each phases\label{IPCMCTMT1}}
\begin{center}
\begin{tabular}{c|c|c|c|c|c}\hline
                 & Disorder  & IOP1     & IOP2     & XY & COP \\ \hline
$\phi_{1}(\pi)   $  & $-$       & $+$      & $+$      & 1  & 2   \\ \hline
$\phi_{1}(\pi /3)$  & $-$       & $-$      & $+$      & 1  & 2   \\ \hline
$\phi_{2}(\pi /3)$  & $-$       & $+$      & $-$      & 1  & $-$   \\ \hline
\end{tabular}
\end{center}
\end{table}

\begin{table}
\caption{Details of the simulations of NERM for the 3D-6GCL model.\label{Hptc_det}}
\begin{center}
\begin{tabular}{c|c|c|c}\hline
lattice size & temperature & Monte Carlo step & the number of samples \\ \hline
60           & 1.383               & 200                      & 20000  \\
             & 1.384               & 200                      & 60000 \\
             & 1.385               & 200                      & 60000 \\
             & 1.386               & 200                      & 20000  \\ \hline
\end{tabular}
\end{center}
\end{table}

\begin{figure}
\caption{Energy level of a neighboring spin pair of the 6GCL model\label{6GCL-1}}
\end{figure}

\begin{figure}
\caption{Order parameter space in 3D-6GCL model. The solid line connecting a vertex and the center is one of six low temperature state and the dotted line between the neighboring solid lines is one of six IOP1 states.\label{6GCL-2}}
\end{figure}

\begin{figure}
\caption{Schematic of the one-spin distribution functions for two kinds of IOP's (a) IOP1 and (b) IOP2.\label{IOP}}
\end{figure}

\begin{figure}
\caption{Three boundary conditions for investigation of the intermediate phase by using the MCTM. (a) $\phi_{1}(\pi)$, (b) $\phi_{1}(\pi /3)$ and (c) $\phi_{2}(\pi /3)$.\label{BC}}
\end{figure}

\begin{figure}
\caption{Temperature dependence of the $\psi$ under the three boundary conditions.\label{IPC_PSI_3BC}}
\end{figure}

\begin{figure}
\caption{Temperature and size dependence of the stiffness exponent under the boundary condition $\phi_{1}(\pi /3)$. \label{IPC_PSI_P3SD}}
\end{figure}

\begin{figure}
\caption{Scaling plots of $-\langle\cos 6\theta\rangle$ for various system sizes and with $\nu |y_{6}|=3.2$ and $c=0.065$, and we use $T_{\mbox{C}}=1.384$ that is estimated by using the NERM. $I_{n}$ is the modified Bessel function.\label{IPC_P6_SC}}
\end{figure}

\begin{figure}
\caption{Snapshot of one of layers under the boundary condition $\phi_2(\pi/3)$ at $T=0.4$.
The colors represent the states. The green cells is the state 1, The red cells is the state 2, and The blue cells is the state 3. There are some of cells of different states by fluctuation. 
\label{profile}}
\end{figure}

\begin{figure}
\caption{Temperature and size dependence of $x$ \label{YOKOYURA}}
\end{figure}

\begin{figure}
\caption{Size dependence of $\chi$ at $T=0.4$ \label{YOKOYURA2}}
\end{figure}

\begin{figure}
\caption{Relaxation of $\lambda_{m}(t)$ \label{Hptc_lm}}
\end{figure}

\begin{figure}
\caption{Relaxation of the $z(t)$ at $T_{\mbox{C}}=1.384$\label{Hptc_Z}}
\end{figure}

\begin{figure}
\caption{Relaxation of the $\beta /\nu (t)$ at $T_{\mbox{C}}=1.384$\label{Hptc_BPN}}
\end{figure}

\begin{figure}
\caption{Scaling plots of $m$\label{Hptc_Beta} for various system size with $\beta =0.34$, and we use $T_{\mbox{C}}=1.384$ and $\beta/\nu=0.515$ that is estimated by using the NERM.  }
\end{figure}

\begin{figure}
\caption{Scaling plots of $\langle\cos 6\theta\rangle$ around the low temperature phase transition with $T_{\mbox{F}}=0.295$\label{LPTC_P6_SC}}
\end{figure}

\begin{figure}
\caption{Temperature and size dependence of the stiffness exponent of $\Delta E$ \label{LPTC_MCTM}}
\end{figure}


\begin{thebibliography}{99}
\bibitem{intro0a}J. L. Cardy, J. Phys. {\bf A 26}, (1982), 6201.
\bibitem{intro0b}D. Nelson, in Phase Transitions and Critical Phenomena, Vol. 7 ed. by C. Domb and J. L. Lebowitz, Academic Press, London. (1983).
\bibitem{intro5}S. Miyashita, J. Phys. Soc. Jpn. {\bf 66} (1997), 3411. 
\bibitem{intro3}Y. Ueno and K. Kasono, Phys. Rev. {\bf B 48} (1993), 16471. 
\bibitem{intro15}A. N. Berker and L. Kadanoff J. Phys. {\bf 13}, (1980), L259.
\bibitem{intro16}F. Matsubara and S. Inawashiro, J. Phys.Soc. Jpn. {\bf 53}, (1984), 4373.
\bibitem{intro6}A. Rosengren and S. Lapinskas, Phys. Rev. Lett. {\bf 71} (1993) 165.
\bibitem{intro7}S. Lapinskas and A.Rosengren, Phys. Rev. {\bf B 49} (1994) 15190.
\bibitem{intro10}R. K. Heilmann, J.-S. Wang and R. H. Swendsen, Phys. Rev. {\bf B 53} (1996) 2210.
\bibitem{intro8}M. Kolesik and M.Suzuki, J. Phys. {\bf A 28} (1995) 6543.
\bibitem{intro9}P. J. Kundrotas, S. Lapinskas, and A. Rosengren, Phys, Rev. {\bf B 52} (1995) 9166.
\bibitem{intro12}O. Koseki and F. Matsubara, J. Phys. Soc. Jpn. {\bf 69} (2000) 1202.
\bibitem{intro4}Y. Ueno, J. Stat. Phys. {\bf 80}, (1995) 841.
\bibitem{intro1}M. Oshikawa, Phys. Rev. {\bf B 61} (2000) 3430. 
\bibitem{intro2}Y. Ueno and R. Kishi, in preparation.
\bibitem{intro14}R. Kishi, Master thesis, Tokyo Institute of Technology (1999).
\bibitem{HPT1}Y. Ozeki ans N. Ito, J. Pys. Soc. Jpn. {\bf 69}, (2000) 193.
\bibitem{HPT2}N. Ito, K. Fukushima, K. Ogawa and Y. Ozeki, J. Pys. Soc. Jpn. {\bf 69}, (2000) 1931.
\bibitem{HPT3}K. Ogawa ahd Y. Ozeki, J. Pys. Soc. Jpn. {\bf 69}, (2000) 2808.
\bibitem{LPT1}T. Takahashi, Master thesis, Tokyo Institute of Technology (1998).
\bibitem{LPT2}T. Takahashi and Y. Ueno, preprint.
\bibitem{DIS1}T. Todoroki, Y. Ueno and F. Matsubara, in preparation.
\bibitem{mekata} M. Mekata, J. Phys. Soc. Jpn. {\bf 42} (1977) 76
\end{thebibliography}
\end{document}